# Application of advanced ultrasonic testing methods to Dissimilar Metal Welds – Comparison of simulated and experimental results

Audrey GARDAHAUT[1], Hugues LOURME[1], Steve MAHAUT[1], Masaki NAGAI[2] and Shan LIN[2].

[1]CEA, LIST, Bât. 565, PC 120, F-91191 Gif-sur-Yvette, France.

[2]Materials Science Research Laboratory, Central Research Institute of Electric Power Industry,
2-6-1 Nagasaka, Yokohama-Shi, Kanagawa-ken 240-0196, Japan.

**ABSTRACT**

Widely present in the primary circuit of Nuclear Power Plants (NPP), Dissimilar Metal Welds (DMW) are inspected using Ultrasonic nondestructive Testing (UT) techniques to ensure the integrity of the structure and detect defects such as Stress Corrosion Cracking (SCC).

In a previous collaborative research, CRIEPI and CEA have worked on the understanding of the propagation of ultrasonic waves in complex materials. Indeed, the ultrasonic propagation can be disturbed due to the anisotropic and inhomogeneous properties of the medium and the interpretation of inspection results can then be difficult. An analytical model, based on a dynamic ray theory, developed by CEA-LIST and implemented in the CIVA software had been used to predict the ultrasonic propagation in a DMW. The model evaluates the ray trajectories, the travel-time and the computation of the amplitude along the ray tube in a medium described thanks to a continuously varying description of its physical properties. In this study, the weld had been described by an analytical law of the crystallographic orientation. The simulated results of the detection of calibrated notches located in the buttering and the weld had been compared with experimental data and had shown a good agreement.

The new collaborative program presented in this paper aims at detecting a real SCC defect located close to the root of the DMW. Thus, simulations have been performed for a DMW described with an analytical law and a smooth cartography of the crystallographic orientation. Furthermore, advanced ultrasonic testing methods have been used to inspect the specimen and detect the real SCC defect. Experimental and simulated results of the mock-up inspection have been compared.

## INTRODUCTION

In worldwide Nuclear Power Plants (NPPs), the welding of stainless steel cooling line pipes and instrumentation components to the ferritic steel vessels are ensured with nickel-based alloys. A large amount of studies have been performed to explain the appearance of leaks in the primary circuit of NPPs [1]-[6]. It has been shown that the formation and the growth of Stress Corrosion Cracking (SCC) are the reason of these leaks and are located in the Dissimilar Metal Welds (DMWs) of the primary circuit. In order to ensure the integrity of the circuit and the reliability of the NPPs, detection and depth sizing of such cracks with high accuracy are essential. To this aim, Ultrasonic Nondestructive Testing (NDT) are commonly used to inspect such welded joints of the primary circuit. However, the interpretation of these on-site controls can be really delicate. Disturbances of the ultrasonic beam such as splitting and skewing [7] appear during the propagation as the materials of the DMW are anisotropic and inhomogeneous [8]. Furthermore, severe scattering and attenuation of the beam, caused by the coarse grain structure of the weld, are also observed. These physical properties of the welds may affect the detection capabilities and the sizing accuracy of the defect. Simulation tools are consequently useful to better understand those phenomena and improve the Ultrasonic Testing (UT) testing applied.

In this context, various models have been developed to simulate the propagation of ultrasonic waves such as finite differences [9], finite elements [10]-[11] or ray-tracing models [12]. In the CIVA software [13]-[14] developed by CEA List, semi-analytical propagation models [15] based on a ray theory have been implemented and applied to the inspection of welds. In this case, the weld is described as a set of several anisotropic homogeneous domains with a given crystallographic orientation determined by observing the macrograph of the weld. For domains smaller than the wavelength, the simulation is however valid only of the contrast of the impedance between two neighboring domains is small [16]16). A new modelling approach has then been considered in order to take into account the inhomogeneity of medium such as DMW. Based on the Dynamic Ray Tracing (DRT) model, this approach is applied on a weld described as a continuously varying description of the crystallographic orientation and allows the evaluation of the ultrasonic propagation in anisotropic inhomogeneous media [17]17).

Therefore, a collaborative research program has been set up between CRIEPI and CEA List to understand the propagation on ultrasonic wave in DMW. Simulations and experiments have then been performed and compared to evaluate the applicability of the CIVA software to the inspection of DMWs with complex internal structures and the phased array techniques for DMWs inspections.

**RESULTS OF THE FIRST COLLABORATION**

In the first collaboration [18]18), CRIEPI and CEA List agreed to study a V-butt dissimilar metal weld made of alloy 600 and connecting SM490 ferritic steel vessels and cooling pipes made of SUS316L stainless steel. The macrograph of the specimen is presented in Figure 1.

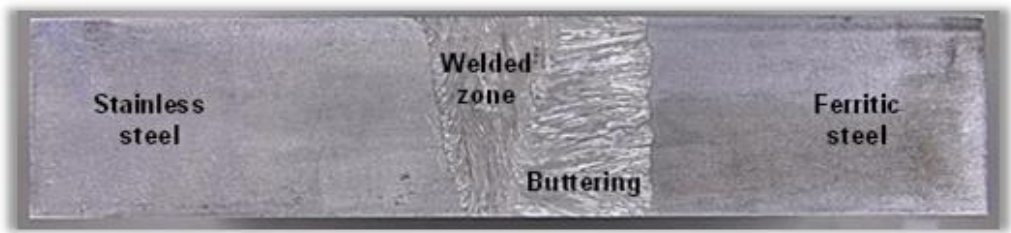

Figure 1 - Macrograph of a V-butt DMW previously considered.

The weld was divided in two distinct domains: the buttering considered as a homogeneous medium with a 78° grain orientation and the weld described as symmetrical with the closed-form expression [12] as presented in Figure 2.

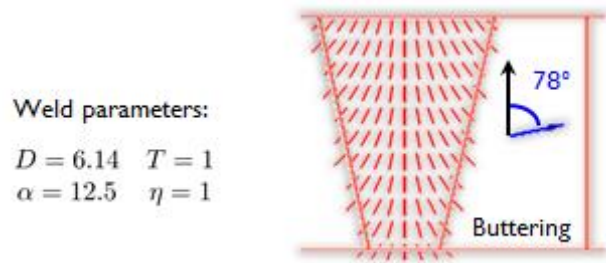

Figure 2 - Representation of the weld orientation in the V-butt DMW of the previous collaboration.

This mock-up contains three 10 mm height notches located in the buttering (S1), in the weld (S2) and in the stainless steel (S3). A 1 MHz linear array transducer of 64 elements (length = 0.5 mm, pitch = 0.6 mm) is used in pulse echo technique and emits a longitudinal wave at 49°. The results of one configuration of control are presented in Figure 3.

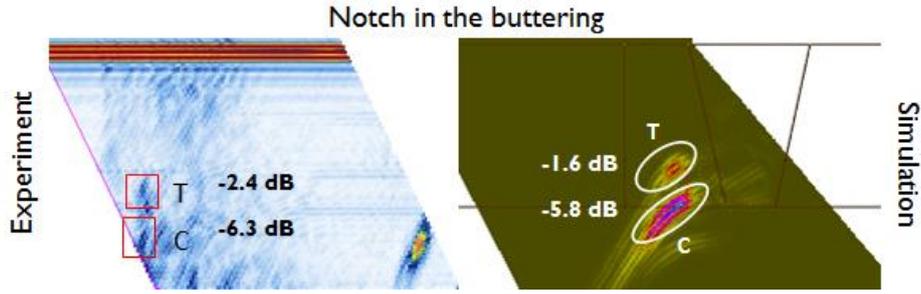

Figure 3 - Experimental and simulated results of the inspection of the buttering defect.

This previous study has highlighted the good agreement between experiments and simulation concerning the detection and the characterization of the EDM notches located in the DMW specimen described with an analytical law.

The objective of this new collaboration was to evaluate the capability of the software to inspect a real SCC located in the root of the weld. Furthermore, a second smooth description of the crystallographic orientation has been used and presented in this paper.

## SIMULATION OF THE ULTRASONIC PROPAGATION WITH A DYNAMIC RAY TRACING MODEL

The Dynamic Ray Tracing model, usually applied in Geophysics 19), is based on the solving of two equations in anisotropic and inhomogeneous media 17).

The first equation, called the eikonal equation, gives the following system called axial ray system 19):

$$\begin{cases} \dfrac{dx_i}{dT} = a_{ijkl} p_l g_j^{(m)} g_k^{(m)} = V_i^{e(m)}, \\ \dfrac{dp_i}{dT} = -\dfrac{1}{2} \dfrac{\partial a_{ijkl}}{\partial x_i} p_k p_n g_j^{(m)} g_l^{(m)}. \end{cases} \quad (1)$$

This system describes the variation of the position $x_i$ and the slowness $p_i$ with respect to the travel-time $T(x)$ and is expressed in function of the elastic constants of the material $a_{ijkl}$, the density $\rho$ and the components of the slowness vector $p_i$. $g_j(m)$ are the eigenvectors of the Christoffel tensor corresponding to the polarization vectors and $V_i^e(m)$ is the energy velocity for the m-mode. Composed of two coupled ordinary differential equations, its solutions give the ray trajectories and the travel-time in the weld.

By solving a second equation, the transport equation, the paraxial ray system is obtained:

$$\begin{cases} \dfrac{d}{dT}\left(\dfrac{dx_i}{d\gamma}\right) = \dfrac{dQ_i}{dT} = \dfrac{1}{2}\dfrac{\partial^2 G}{\partial p_i^{(x)}\partial\gamma}, \\ \dfrac{d}{dT}\left(\dfrac{dp_i^{(x)}}{d\gamma}\right) = \dfrac{dP_i}{dT} = \dfrac{1}{2}\dfrac{\partial^2 G}{\partial x_i \partial\gamma}. \end{cases} \quad (2)$$

Composed of ordinary linear differential equations of the first order of the paraxial quantities $Q_i$ and $P_i$, it expresses the variations of the position $x_i$ and the slowness $p_i$ in function of the travel-time $T(x)$ and the parameter $\gamma$ which is any parameter of a ray $\Omega$ and is chosen as a take-off angle between a reference axis and the initial slowness vector. Gm, the normalized eigenvalues of the Christoffel tensor, are expressed as $G_m = a_{ijkl} p_j p_l g_i^{(m)} g_k^{(m)}$. By solving this second system, the evolution of the ray tube during the propagation is obtained and its amplitude, assuming the conservation of the energy across a ray tube section, is evaluated.

These two systems are solved simultaneously by using numerical techniques such as the Euler method in this example.

## PRESENTATION AND DESCRIPTION OF THE DMW TEST BLOCK

The mock-up of this second collaboration connects a SM490 ferritic steel pipe to a SUS316L stainless steel pipe. It is composed of a V-butt dissimilar metal weld made of alloy 600, a buttering made with alloy 82 and a SUS cladding as shown in Figure 4.

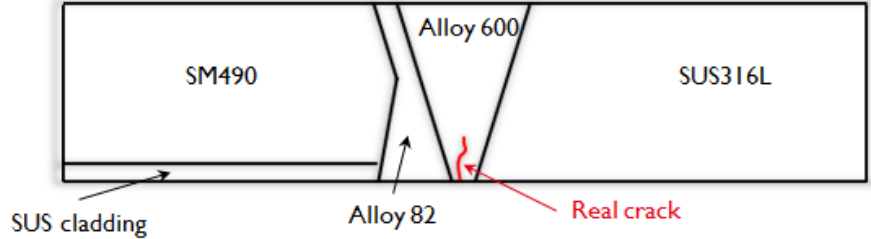

Figure 4 - Scheme of the DMW test block.

This specimen contains a real stress corrosion cracking located in the middle of the weld root. A micrograph of the defect is presented in Figure 5. As shown in the figure, the dimension of the SCC is different along the weld bead. It is quite vertical with a height of 3.1 mm on one side and tilted with a height of 2.3 mm on the other side.

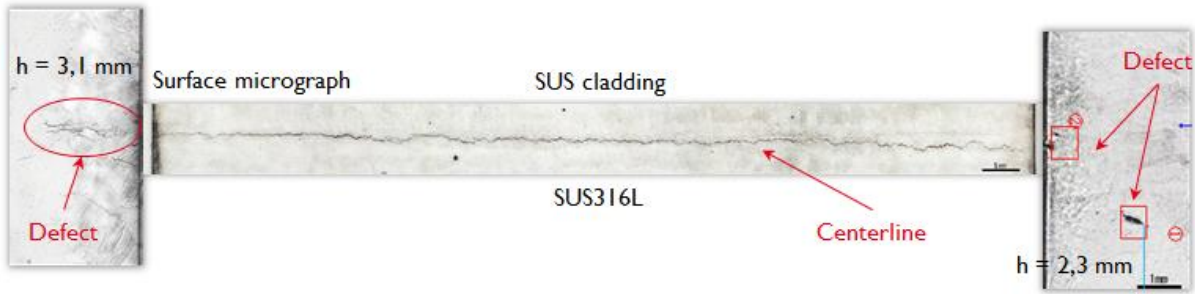

Figure 5 - Micrographs of the inner surface and the two sides of the specimen.

As said previously, DMWs are characterized by their anisotropic and inhomogeneous properties and a strong attenuation. These data are used as input data in the CIVA software. The anisotropy of the material results of the dependence of the ultrasonic velocity to the direction of propagation. It is expressed through the elastic constants. The chosen set of parameters is presented in Table 1. These properties of alloy 182 are representative of the anisotropy of the studied V-butt weld made of alloy 600.

|  | $C_{11}$ | $C_{22}$ | $C_{33}$ | $C_{23}$ | $C_{13}$ | $C_{12}$ | $C_{44}$ | $C_{55}$ | $C_{66}$ | $\rho$ |
|---|---|---|---|---|---|---|---|---|---|---|
| Alloy 182 [19] | 236.1 | 255.8 | 255.8 | 130.5 | 137.9 | 135.4 | 81.4 | 111.4 | 111.9 | 8260 |

Table 1 - Input parameters in CIVA for alloy 182. Elastic constants in GPa and density in kg.m$^{-3}$.

During the propagation, the attenuation is caused by the wave scattering on the constitutive macroscopic grains and the absorption linked to the material viscosity. This energy loss has been evaluated at 2 MHz in the testing block. For the buttering the attenuation is 0.165 dB.mm$^{-1}$ while it is 0.292 dB.mm$^{-1}$ in the welded zone.

The DRT model relies on a high frequency approximation meaning the weld has to be described as a continuously varying description of the grain orientation. Two description have been used in this study to express the inhomogeneity of the medium. The first one is an analytical law [12] expressed as:

$$\theta = \begin{cases} \arctan(\dfrac{T(D+z\tan\alpha)}{x^\eta}), & \text{for } x > 0, \\ -\pi/2, & \text{for } x = 0, \\ -\arctan(\dfrac{T(D+z\tan\alpha)}{(-x)^\eta}), & \text{for } x < 0. \end{cases} \quad (3)$$

$T$ and $\eta$ represent the evolution of the grain orientation and $D$ and $\alpha$ express the geometry of the V-butt weld. The weld zone is then described smoothly with $D = 4.25$ mm, $\alpha = 17.8°$, $T = 0.75$, $\eta = 1$ while the buttering is considered as a homogeneous domain with a 90° crystallographic orientation. The second description is obtained by applying an image processing technique on the weld macrograph 21). Both descriptions of the weld are presented in Figure 6.

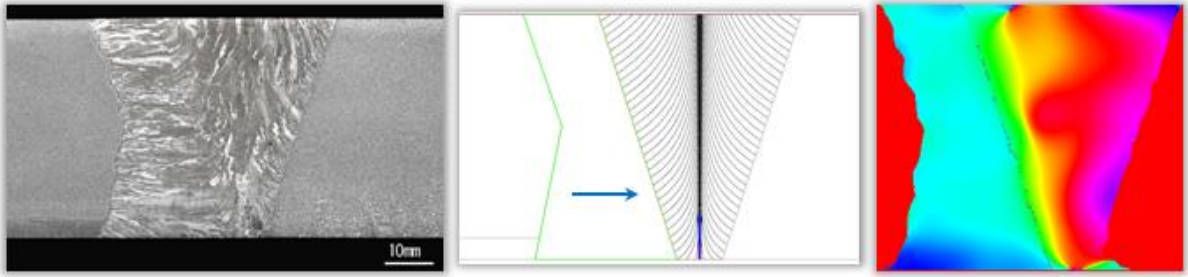

Figure 6 - Representation of the macrograph (left), the analytical description (center) and the smooth cartography of the crystallographic orientation (right) of the DMW.

**PRESENTATION OF THE EXPERIMENTAL SETUP**

The experiments performed in the framework of this study aim at detecting and sizing, when it is possible, the real stress corrosion cracking located at the inner surface in the weld root. A pulse-echo technique has been used with linear phased array probes emitting longitudinal waves at 1 and 2 MHz and positioned in the outer surface of the specimen. The active surface of the transducer is composed of 64 elements with 0.5x20 mm dimensions and the pitch is 0.6 mm. It has been chosen to inspect the specimen from both sides of the weld in order to maximize the probability of detection and characterization of the defect. As presented in Figure 7, the probes have been move mechanically in the two directions.

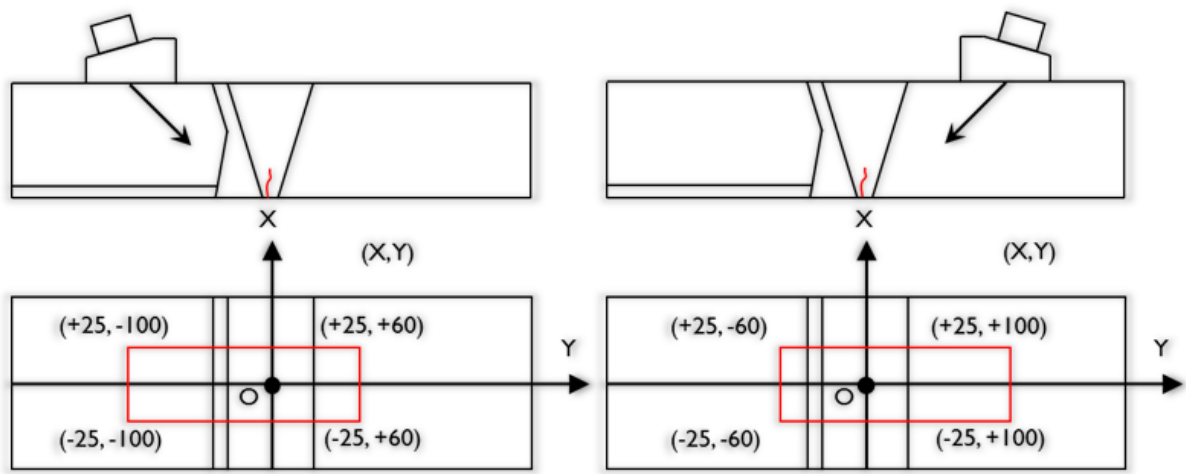

Figure 7 - Setup of the experimental campaign. Position and displacement of the probes.

The experimental results of the inspection performed with a 1 MHz linear phased array probe are presented in Figure 8 while the results obtained with a 2 MHz linear phased array transducer are observed in Figure 9.

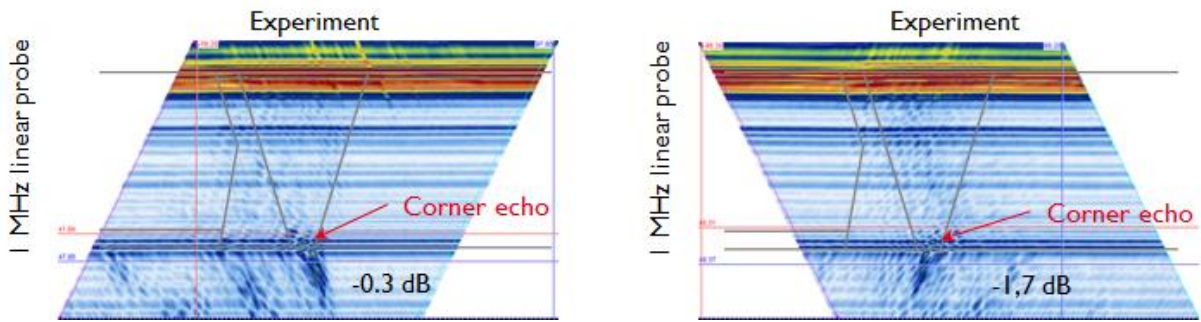

Figure 8 - Experimental results for a 1 MHz linear phased array probe. The inspection has been performed from the stainless steel (left) and the ferritic steel (right).

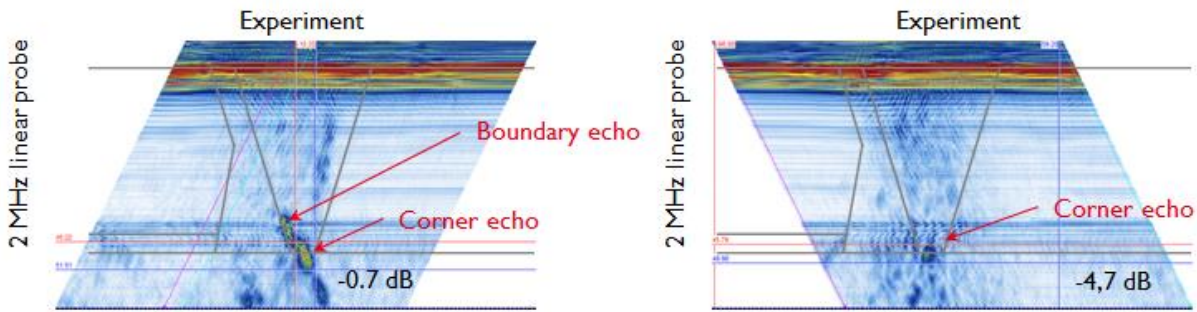

Figure 9 - Experimental results for a 2 MHz linear phased array probe. The inspection has been performed from the stainless steel (left) and the ferritic steel (right).

It is presented in these results that the corner echo of the real stress corrosion cracking is well observed at 1 and 2 MHz but the tip diffraction echo of the top of the defect is not detected. This results was expected as the defect is smaller than the wavelength at 1 MHz so the tip diffraction cannot be detected. At 2 MHz its maximal height is around the wavelength so its response could be observed at this specific position but cannot be ensured as the characteristics of the defect are not identical all along the weld bead. Furthermore, it can be observed that the amplitude of the defect is lower when the ultrasonic wave has to propagate through the buttering and the weld regions so the detection from the ferritic steel side is less favorable.

As presented previously, the associated simulation have been performed with CIVA for a weld described thanks to an analytical law and with a smooth cartography of the crystallographic orientation. The defect has been chosen as a vertical notch with a 3.1 mm height. The results obtained at 1 and 2 MHz are presented in Figure 10 and Figure 11 respectively.

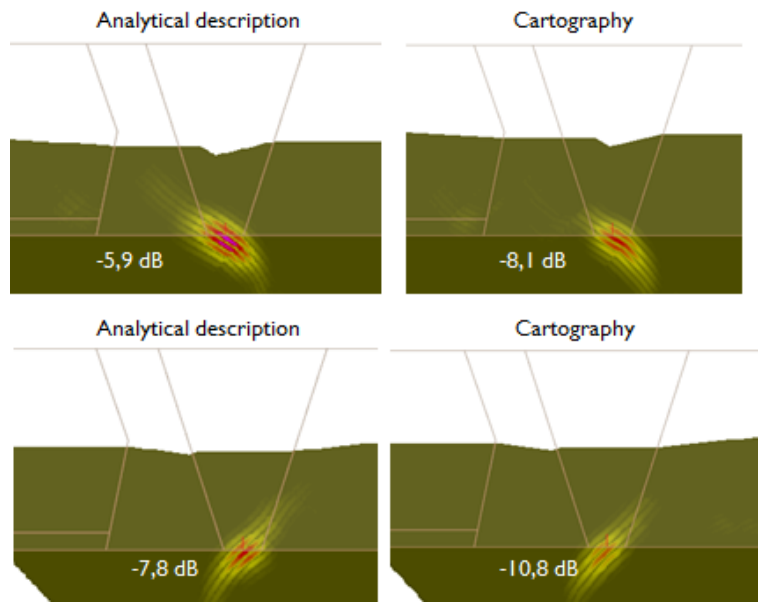

Figure 10 – Inspection simulation results for a 1 MHz linear phased array probe. The inspection has been performed from the stainless steel (top) and the ferritic steel (bottom).

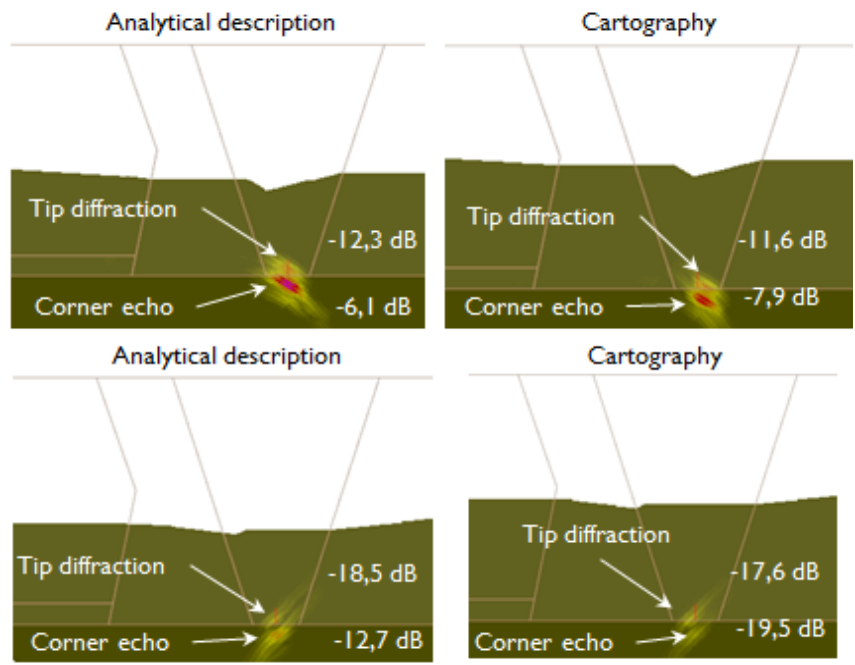

Figure 11 - Inspection simulation results for a 2 MHz linear phased array probe. The inspection has been performed from the stainless steel (top) and the ferritic steel (bottom).

In simulation, the corner echo is reproduced and the tip diffraction of the top of the notch is also observed at 2 MHz. As explained for the experimental results this result was expected. Indeed, the defect is smaller than the wavelength at 1 MHz so the tip diffraction cannot be simulated. However, at 2 MHz the defect height height is around the wavelength. The resolution is then better so the defect response is simulated.

Concerning the amplitude of the defect, a good agreement is observed between the inspection simulation results obtained for both descriptions of the crystallographic orientation. Furthermore, as for the experimental results it can be seen that the amplitude of the defect echoes are higher when the propagation is made only in the weld. The second path, from the ferritic steel is indeed less favorable as two media with different physical properties are crossed.

However, some discrepancies can be noted between the experimental and simulated results around 6 and 8 dB. The defect inspected during the experiments is a real stress corrosion cracking located in the middle of the weld root while the first step of the simulation aimed at evaluating the response of a vertical notch. Indeed the shape of the real crack and its tilt were not taken into account. As presented in Figure 12, the inclination of the defect has a significant influence on the amplitude of the response. In this configuration indeed, the variation of 10 of the defect angle comes out at a variation of 3.5 dB for the amplitude of the corner echo.

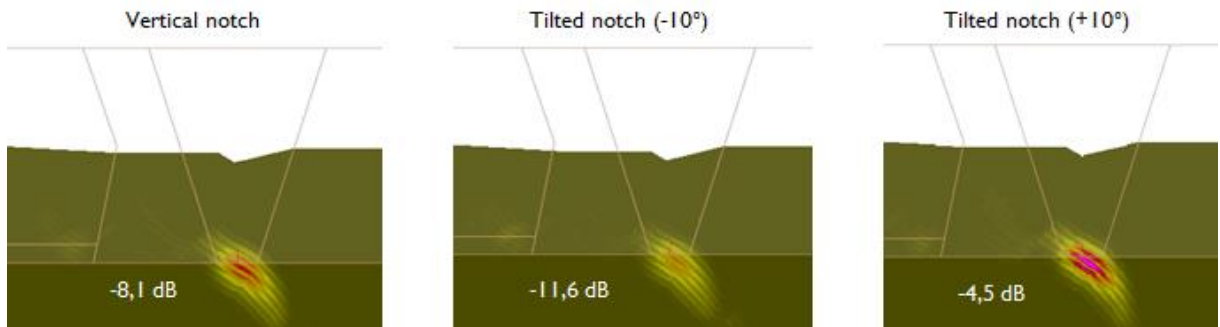

Figure 12 – Illustration of the influence of the tilt on the amplitude of the defect.

More, the attenuation values have been obtained for an only angle of 0° between the incident beam and the grain orientation. Multiple studies have highlighted the influence of this parameter on the propagation and particularly the ultrasonic attenuation. Concerning the geometry of the welded region, the border of the buttering has been considered as straight lines in simulation in order to avoid some artefacts due to the reaching of the model limits. The interfaces have also to be smooth to respect the model limitation so the description is not accurate compared to the macrograph. Finally, the experimental results have been chosen at the middle of the specimen width where we do not have any information about the defect shape and inclination. The amplitude of the response would be better evaluated with this knowledge.

**CONCLUSION AND PERSPECTIVES**

This paper has presented the experimental and simulated results of the ultrasonic inspection of a V-butt dissimilar metal weld made of alloy 600 and containing a real stress corrosion cracking in its root. The purpose was to evaluate the capability of the DRT model implemented in the CIVA software to compute the ultrasonic propagation in anisotropic and inhomogeneous media. In a previous study, the weld has been described with an analytical law from the observation of the macrograph and an EDM notch has been implanted in the specimen and inspected. In this work, the crystallographic orientation has been described with an analytical law anew but also with a smooth cartography. More, the defect considered has been a real inner surface-breaking stress corrosion cracking located in the middle of the weld root.

A good agreement has been observed between the simulation and experiments about the detection of the corner echo of the real SCC. Indeed, this echo has been observed for every configuration. It has been shown that the tip diffraction echo has been simulated at 2 MHz as expected but has not been observed in experiments which can be explained by the lack of homogeneity of the defect shape and characteristics along the weld bead and the influence of the internal structure on the ultrasonic propagation.

However, discrepancies of the amplitude of the defect response have been highlighted between the experiments and the simulations. A destructive testing has been considered in order to evaluate the crack shape by fracture observation at different positions in the specimen. This information will then be taken into consideration in simulations. First a study will be performed to evaluate the influence of the inclination of the defect on its response and more specifically on its amplitude and second, a multi-faceted defect will be simulated in the CIVA software. The variation of the defect response along the weld bead will also been considered in order quantify the variation of the amplitude and compare simulations and experiments. Finally, it is considered to apply other advances ultrasonic techniques to the inspection of the real stress corrosion cracking located in a weld root such as FMC/TFM method [22] and Plane Wave Imaging (PWI) technique [23].

**REFERENCES**

1) BRAATZ B G *et al.*, "Primary Water Stress Corrosion Cracks in Nickel Alloy Dissimilar Metal Welds: Detection and Sizing Using Established and Emerging Nondestructive Examination Techniques", IAEA-CN-194-025, pp.1-9 (2012).

2) BAMFORD W H *et al.*, "Integrity Evaluation for Future Operation: Virgil C. Summer Nuclear Power Plant Reactor Vessel Nozzle to Pipe Weld Regions", WCAP-15615, Rev. 0, Westinghouse Engineering (2000).

3) BAMFORD W H *et al.*, "Alloy 182 Crack Growth and its Impact on Service-Induced Cracking in PWR Plant Piping", (*Proc. 10$^{th}$ International Conference on Environmental Degradation of Materials in Nuclear Power Systems – Water Reactors, Houston, Texas*), NACE, Paper 34 (2002).

4) JENSSEN A *et al.*, "Assessment of Cracking in Dissimilar Metal Welds", (Proc. 10$^{th}$ International Conference on Environmental Degradation of Materials in Nuclear Power Systems – Water Reactors, Houston, Texas), NACE (2002).

5) JENSSEN A *et al.*, "Structural Assessment of Defected Nozzle to Safe-End Welds in Ringhals-3 and -4", (Proc. Fontevraud V International Symposium), SFEN (2000).

6) CUMBLIDGE S E *et al.*, "Nondestructive and Destructive Examination Studies on Removed from-Service Control Rod Drive Mechanism Penetrations", NUREG/CR-6996, PNNL-18372, U.S. Nuclear Regulatory Commission, Washington, D.C. (2009).

7) CHASSIGNOLE B *et al.*, "Ultrasonic Examination of Austenitic Weld: Illustration of the Disturbances of the Ultrasonic Beam", *35$^{th}$ Review of Progress in Quantitative Nondestructive Evaluation*, Vol. 28, pp. 1886-1893 (2009).

8) KUPPERMAN D S and REIMANN K J, "Ultrasonic Wave Propagation and Anisotropy in Austenitic Stainless Steel Weld Metal", *IEEE Transactions on Sonics and Ultrasonics*, Vol. SU-27, No. 1, pp. 7-15 (1980).

9) FELLINGER P *et al.*, "Numerical Modeling of Elastic Wave Propagation and Scattering with EFIT – Elastodynamic Finite Integration Technique", *Wave Motion*, Vol. 21, pp. 47-66 (1995).

10) APFEL A *et al.*, "Coupling an Ultrasonic Propagation Code with a Model of the Heterogeneity of Multipass Welds to Simulate Ultrasonic Testing", *Ultrasonics*, Vol. 43, pp. 447-456 (2005).

11) CHASSIGNOLE B *et al.*, "Modelling the Attenuation in the ATHENA Finite Elements Code for the Ultrasonic Testing of Austenitic Stainless Steel Welds", *Ultrasonics*, Vol. 49, pp. 653-658 (2009).

12) OGILVY J A, "Computerized ultrasonic Ray Tracing in Austenitic Steel", *NDT International*, Vol. 18, No. 2, pp. 67-77 (1985).

13) CIVA software platform for simulating NDT techniques (UT, EC, RT) http://www-civa.cea.fr.

14) LHÉMERY A *et al.*, "Modeling Tools for Ultrasonic Inspection of Welds", *NDT&E International*, Vol. 37, pp. 499-513 (2000).

15) GENGEMBRE N and LHÉMERY A, "Pencil Method in Elastodynamics: Application to Ultrasonic

Field Computation", *Ultrasonics*, Vol. 38, pp. 495-499 (2000).

16) GARDAHAUT A *et al.*, "Evaluation of Ray-Based Methods for the Simulation of UT Welds Inspection", *AIP Conference Proceedings*, Vol. 1511, pp. 1073-1080 (2013).

17) GARDAHAUT A *et al.*, "Paraxial Ray-Tracing Approach for the Simulation of Ultrasonic Inspection of Welds", *AIP Conference Proceedings*, Vol. 1581, pp. 529-536 (2014).

18) GARDAHAUT A *et al.*, "Application of a 3D ray tracing model to the study of ultrasonic wave propagation in Dissimilar Metal Welds", *11th International Conference on NDE in Relation to Structural Integrity for Nuclear and Pressurized Components*, 19-21 May, Jeju, Korea (2015).

19) ČERVENÝ V, *Seismic Ray Theory*, Cambridge: Cambridge University Press (2001).

20) CHASSIGNOLE B *et al.*, "Ultrasonic and Metallurgical Examination of an alloy 182 Welding Mold", *7th International Conference on NDE in Relation to Structural Integrity of Nuclear and Pressurized Components*, 12-15 May, Yokohama, Japan (2009).

21) GARDAHAUT A, "Développement d'outils de modélisation pour la propagation ultrasonore dans les soudures bimétalliques", PhD Thesis, Université Paris Diderot – Paris 7 (2013).

22) FIDAHOUSSEN A *et al.*, "Imaging of defects in several complex configurations by simulation-helped processing of ultrasonic array data", *AIP Conference Proceedings*, Vol. 1211, pp. 847-854 (2010).

23) LE JEUNE L *et al.*, "Plane Wave Imaging for ultrasonic non-destructive testing: Generalization to multimodal imaging", *Ultrasonics*, Vol. 64, pp. 128-138 (2015).